\documentclass[12pt]{article}
\usepackage[russian]{babel}
\begin{document}
\begin{center} {\bf Topological Interpretation of the Dirac Equation and Geometrical Foundations of the Gauge Invariance Idea}\end{center}
\begin{center}  О.А. Ol'khov* \end{center}
\begin{center}{\it Institute of Chemical Physics, Russian Academy of Sciences, 119991, Moscow, Kosygin Str.,4, Russia}\end{center}

Soon after the Yang-Mills work, the gauge invariance became one of
the basic principles in the elementary particles theory. The gauge
invariance idea is that Lagrangian has to be invariant not only with
respect to the coordinates transformations corresponding to the
Lorentz group (external symmetry). It is supposed that Lagrangian
has also to be invariant with respect to wave functions (not
coordinates) transformations corresponding to some additional groups
(so-called "internal symmetry groups"). Useful though this idea is,
there is no satisfactory understanding of the above additional
symmetries origin, and the gauge invariance is considered as an
auxiliary theoretical hypotheses. We propose a new, topological
interpretation of the basic quantum mechanical equation --- the
Dirac equation, and within the framework of this interpretation the
notions of internal symmetry and gauge invariance bear a simple
geometrical meaning and are natural consequences of the basic
principles of the proposed geometrical description. According to
this interpretation the Dirac equation proves to be the
group-theoretical relation that accounts for the symmetry properties
of a specific 4-manifold --- localized microscopic deviation of the
space-time geometry from the Euclidean one. This manifold covering
space plays the role of the internal space, and the covering space
automorphism group plays the role of the gauge group.

\begin{center}{\bf 1. Introduction} \end{center}

At high energies, the quantum mechanics formalism is now considered
as a description of interacting quanta of different physical wave
fields: electron--positron field, quark field, etc. At low energies,
where the probability of new particles appearance is negligible,
above formalism is considered as a description of interacting
pointlike particles having some wave properties (wave--corpuscular
duality concept). In this presentation we show that the Dirac
equation can be considered as a relationship between some geometry
and group-theory notions. In result, this relationship can be
considered as a description of symmetry proprieties of some specific
microscopic space--time deformation. The macroscopic deformation of
the space--time is the basic idea of a geometrical interpretation of
the gravitational field within the framework of the general theory
of relativity. So, the suggested interpretation, in some sense, is
an attempt to extend the general relativity ideology to the
description of microscopic quantum objects.

There is another aspect of this contribution. There were many
attempts to find out some new interpretation of quantum mechanics
formalism. On the other hand, in parallel, there were many attempts
to geometrize electromagnetic and other physical fields. The
distinctive feature of this work is that it combines both above
problems within one idea. Namely, we show that the equations for the
gauge electromagnetic field and the equations for this field sources
can be considered as a description of a single geometrical object.
Some preliminary results were published earlier [1-5].

This paper is organized as follows. In Section 2 we show that the
Dirac equation for a free particle can be interpreted as a
group-theoretical description  of the specific space--time
deformation, namely, the closed nonorientable four-dimensional
manifold (4-manifold). Differentiation operators and Dirac's
matrices in this equation prove to be some symmetry-group operators
and the wave function proves to be a basis vector of the
corresponding symmetry-group representation. In Section 3 we extend
the above interpretation to the Dirac equation for a charged
particle interacting with its own electromagnetic field. Within
geometrical interpretation, this equation happens to be a
group-theoretical relation written for some curved 4-space that is a
standard  geometrical characteristic of a closed 4-manifold
(universal covering space of the manifold). This space proves to be
the analogy of so called particle "internal" space --- a notion
additionally introduced into the relativistic quantum mechanics. So,
within the geometrical interpretation, the notions of internal space
and its internal symmetry are not some additional attributes of the
theory but are natural consequences of the geometrical description.
Section 4 contains conclusion and final remarks. In Appendix, we
illustrate by the simplest example (one-dimensional closed manifold)
the possibility of the manifold description by means of linear
differential equations.

\begin{center}{\bf 2. Equation for free particle}\end{center}

Let us first consider the Dirac equation for a free particle [6],
$$i\gamma_{\mu}\partial_{\mu}\psi=m\psi, \eqno(1)$$ where
$\partial_{\mu}=\partial/\partial x_{\mu},\quad \mu =1,2,3,4,\quad
\gamma_{\mu}$ are the Dirac matrices, $\psi (x)$ is the Dirac
bispinor, $x_{1}=t, x_{2}=x, x_{3}=y$, and $x_{4}=z$. There is a
summation over repeating indices with a signature $(1,-1,-1,-1)$.
Here, $\hbar =c=1$. The solution to Eq.(1) has the form of a plane
wave
$$\psi=u\exp (-ip_{\mu}x_{\mu}),\eqno(2)$$ where the 4-momentum $p_{\mu}$
satisfies the relation
$$p_{1}^{2}-p_{2}^{2}-p_{3}^{2}-p_{4}^{2}=m^{2}.\eqno(3)$$

The interpretation of Eq.(1) is, in fact, defined by the
interpretation of its solution (2). The up-to-date interpretation
depends on the particle energy. For $c\rightarrow\infty$
(nonrelativistic limit), $|\psi (x)|^2$ is interpreted as a
probability to detect the particle at point $x$. The wave function
$\psi (x)$ in (2) is said to represent, in this case, the
"probability traveling wave" (the wave-corpuscular duality concept).
For $v\sim c$, the probability of the new particles appearance needs
to be taken into account. In this case, the wave function is
interpreted as the annihilation and creation operator for the
corresponding field quantum.

Let us show that the geometrical and symmetry characteristics of a
certain closed space--time 4-manifold are encoded in Eqs.(1--3) by
means of the group-theory operators. It is now be helpful to rewrite
Eqs.(1--3) in a more convenient form. We rewrite Eq.(1) as
$$
(l_{1}^{-1}\gamma_{1}T_{1}-l_{2}^{-1}\gamma_{2}T_{2}-l_{3}^{-1}\gamma_{3}T_{3}-l_{4}^{-1}\gamma_{4}T_{4})\psi
=l_{m}^{-1}\psi,\eqno(4)$$ where operator $T_{\mu} (\mu =1,2,3,4)$
has the form
$$T_{\mu}=i(2\pi)^{-1}l_{\mu}\partial_{\mu},\quad
l_{\mu}=2\pi p_{\mu}^{-1},\quad l_{m}=2\pi m^{-1}.\eqno(5)$$ We
rewrite the solution (2) as
$$\psi=u\exp (-2\pi
ix_{1}l_{1}^{-1}+2\pi ix_{2}l_{2}^{-1}+2\pi ix_{3}l_{3}^{-1}+2\pi
ix_{4}l_{4}^{-1}).\eqno(6)$$ We also rewrite the conservation law
(3) as
$$l_{1}^{-2}-l_{2}^{-2}-l_{3}^{-2}-l_{4}^{-2}=l_{m}^{-2}.\eqno(7)$$
Note that all quantities in Eqs (4--7) has the dimensionality of
length.

Operators $T_{\mu}$ in Eq.(5) and functions $\psi (x)$ in Eq.(6) are
related by the following equation.
$$\psi^{'}(x^{'}_{\mu})\equiv T_{\mu}\psi(x_{\mu}^{'})=\psi(x_{\mu}),\quad
x_{\mu}^{'}=x_{\mu}+l_{\mu}.\eqno(8)$$ Within a group theory the
relation (8) means that $T_{\mu}$ is the operator representation of
a discrete group of parallel translations along $x_{\mu}$ axis with
the generator $l_{\mu}$. The relation (8) means also that the
function $\psi (x)$ is a basic vector of this representation [7].

In addition, being a four-component spinor, the function $\psi (x)$
is related to the four-row Dirac matrices $\gamma_{\mu}$ by the
equations [8,9]
$$\psi^{'}(x^{'})=\gamma_{\mu}\psi (x),\eqno(9)$$ where $x\equiv
(x_1,x_2,x_3,x_4)$,  $x^{'}\equiv (x_1,-x_2,-x_3,-x_4)$ for $\mu
=1$, and $x^{'}\equiv (-x_1,x_2,-x_3,-x_4)$ for  $\mu =2$, and so
on. Within the group theory, Eqs.(9) mean that the matrices
$\gamma_{\mu}$ are a matrix representation of the group of
reflections along three axes perpendicular to the $x_{\mu}$ axis.
Relation (9) means also that the function $\psi (x)$ is a basic
vector of the above symmetry-group representation.

A parallel translation with simultaneous reflection in the direction
perpendicular to the translation is often spoken of as "sliding
symmetry" [10]. Thus, we see that the operators
$$P_{\mu}=\gamma_{\mu}T_{\mu}.\eqno(10)$$
form the representations of sliding symmetry group (sliding symmetry
in the direction $0x_{\mu}$). Using the above notation, we can
rewrite Eq.(1) as
$$l_{\mu}^{-1}P_{\mu}\psi=l_{m}^{-1}\psi,\eqno(11)$$ where, as we
established before, the function $\psi (x)$ can be interpreted as a
basic vector of the above sliding symmetry group that operates in
the four-dimensional space, in which Eqs.(1) and (11) are written.

There is an important moment of the presentation at this point.
Eq.(1) is usually supposed to be written in the four-dimensional
space--time that has no any sliding symmetry. The physical
space--time has a symmetry only with respect to translations and
4-rotations (Poincar\'{e} group). This symmetry reflects the
space--time homogeneity and, therefore, the Dirac equation (1) is an
invariant with respect to this symmetry group [6]. What does the
appearance of a new additional symmetry means? The suggestion about
the reasons for an appearance, within mathematical formalism, of a
new additional symmetry is a basis of the proposed geometrical
interpretation of the Dirac equation (1).

Our main suggestion is that the Dirac equation (1) describes some
special microscopic distortion of the physical space--time, and it
is necessary to have some auxiliary four-dimensional space with the
sliding symmetry to represent the distortion geometrical proprieties
mathematically. The objects of this kind are known in geometry.
These are so called closed topological manifolds. Such manifold is
defined  as any of the objects that can be obtained from some
definite geometrical object by its deformation without
discontinuities [10]. The above mentioned auxiliary space is spoken
of as the manifold universal covering space. The symmetry group of
this space coincides with the so-called fundamental group of the
closed manifold. This group is the group of different classes of
closed paths starting and ending at the same point of the manifold
and this group is one of the important topological characteristics
of manifolds.

Thus, our suggestion is that the "discovered" sliding symmetries
along four axes  are symmetries of the four-dimensional universal
covering space of the closed space--time manifold. It means that the
fundamental group of this manifold is a group of four sliding
symmetries. These four sliding symmetries correspond to four
possible classes of closed paths. The Dirac equation (1) is supposed
to be written for the manifold covering space, and Eq.(11) has to be
considered as the group-theoretical relation that leads to the
restriction (7) for the four possible values of $l_{\mu}$ ($l_{\mu}$
is a length of the closed path for one of the four possible
classes).

By now, the topology of four-dimensional manifolds is understood to
a much lesser extent than the topology of, e.g, two-dimensional
manifolds. Only two-dimensional manifolds are classified in detail,
and the parameters of their symmetry groups are known [10]. For this
reason, we have no opportunity to identify our space--time
4-manifold with any well investigated topological object. So, we
will attempt to invoke a possible analogy with some two-dimensional
manifold having formally almost the same proprieties as the
four-dimensional space-time manifold that we introduced before. The
fundamental group of this two-dimensional manifold is generated by
two sliding symmetries, and its universal covering space is a
two-dimensional Euclidean space. We have in mind the so-called Klein
bottle. The Klein bottle is one of the simplest examples of
nonorientable geometrical objects and it can be obtained by gluing
together two nonorientable M\"{o}bius strips [10]. So we can suggest
that Eq.(11) describes microscopic closed nonorientable
four-dimensional space-time manifold similar formally to the
two-dimensional Klein bootle. Microscopic sizes of this manifold are
defined by microscopic values of its length parameters $l_{\mu}$ and
$l_{m}$. For example, $l_{m}=2\pi \hbar /mc\sim 10^{-10}$ cm., if
$m$ is the electron mass.

\begin{center}\textbf{3. Geometrical interpretation of the notions of "internal
space", "internal symmetry", and "gauge invariance"}\end{center}

There is, of course, a question whether the suggested geometrical
interpretation of Eg.(1) has any advantages in compare with the
traditional one. We will show below that the geometrical
interpretation resolves one of the problems of the relativistic
quantum mechanics that has yet no solution within the framework of
traditional interpretation.

One of the basic idea of the relativistic quantum mechanics is the
suggestion that the basic equations have to be invariant with
respect to some additional "internal" or "gauge" symmetry groups
that differ from the "external" Poincar\'{e} group and that operate
in some  additional "internal space" of a particle. This suggestion
is usefully employed in the gauge theory of elementary particles but
there is no satisfactory explanation of these additional symmetries
[11,12]. We will show below that, within the framework of the
geometrical interpretation, the notions of internal symmetry,
internal space, and gauge invariance bear a simple geometrical
meaning and are natural consequences of the basic principles of
geometrical description. We have seen in the previous section that
the geometrical interpretation of Eq.(1) introduces into
consideration the additional space---the universal covering space of
4-manifold. This fact will serve as a guide for geometrical
interpretation of the "internal space."

We will consider the problem for a case of the classical Dirac
electron--positron field interacting with its own gauge
electromagnetic field. The system of equations for this case (Dirac
and Maxwell equations) has the form [6]
$$i\gamma_{\mu}(\partial_{\mu}-ieA_{\mu})\psi=m\psi.\eqno (12)$$
$$F_{ik}=\partial_{k} A_i-\partial_{i} A_k,\eqno (13)$$
$$\partial_{i} F_{ik}=j_{k}.\eqno (14)$$ Here,
$e$ and $m$ are electron charge and mass, respectively, $A_l$ are
4-potentials of electromagnetic fields, and the current $j_{k}$ is
defined by the relation
$$j_{k}=e\psi^{+}\gamma_{k}\psi,\eqno (15)$$ where
$\psi^{+}=\psi^{*}\gamma_{1}$.

The only distinction of Eq.(12) from Eq.(1) is the replacement of
conventional derivatives $\partial_{\mu}$ by so-called long
derivatives ($\partial_{\mu}-ieA_{\mu}$). Inserting electromagnetic
4-potentials in Eq.(1) by means of the above-mentioned replacement
is a consequence of the suggestion that Eq.(12) is the invariant
with respect to the phase transformation of the wave function
[11,12]
$$\psi \rightarrow \psi^{'}=\psi \exp [ie\alpha (x)],\eqno (16)$$ where $\alpha
(x)$ is a parameter depending on the space-time coordinate $x$. This
invariance (named as "gauge invariance")  may be realized only by
inserting electromagnetic 4-potentials into Eq.(1) as it is done in
Eq.(12). In so doing, we obtain the following gauge relation for
$A_{l}$ [11,12].
$$A_l \rightarrow  A_l^{'}=A_l+\partial_{l} \alpha(x).\eqno
(17)$$ (Note that the above transformation leaves $F_{ik}$
unchanged). Relation (16) does not directly include a space-time
(external) coordinates transformation, and this fact is said to be
the reflection of some "internal space" symmetry. As we mentioned
before, the nature of this internal space and the origin of its
gauge symmetry have no satisfactory explanation [11].

Let us now look at requirements of the gauge invariance (16,17)
within the framework of the possible geometrical interpretation of
Eq.(12). Within the geometrical approach, we have to consider
Eq.(12) (similarly to Eq. (1)) as a group-theoretical relation for a
four-dimensional (may be non-Euclidean) universal covering space of
some space--time 4-manifold. But it will be the case only if we
could prove that the long derivatives in Eq.(12) can be considered
within a differential geometry as so-called covariant derivatives
for some non-Euclidean space. Moreover, we have to find out
transformations of this space coordinates, that correspond to a
gauge transformation of wave function (16), and we must prove that
these coordinate transformations lead to the gauge relation (17).

It happened that the above-mentioned mathematical problem was long
ago solved by Weil, who considered a different physical problem
[15]. Weil has attempted to represent an electromagnetic field
(analogously to a gravitational field) as a manifestation of the
physical space-time curvature. He has proved, that the long
derivatives in (12) $(\partial_{l}-ieA_{l})$ can be, indeed,
considered as covariant derivatives for the special curved 4-space
(the space with semimetrical parallel translation). In this case,
the values $ieA_{l}$ play a role of the space specific geometrical
characteristic---"connectivity". What is more, Weil has showed that
gauge transformations (16,17) may be considered as a consequence of
the following transformation of the space-time metric $g_{ik}$:
$$g_{ik}^{'}=\lambda (x) g_{ik}.\eqno (18)$$ The infinitesimal element of the 4-space length (the 4-interval $ds$) is defined by the
relation [13]
$$ds^{2}=g_{ik}^{'}dx_{i}dx_{k}.$$ Therefore,
transformation (18) corresponds to the 4-space isotropic stretching
depending on space coordinates.

The real physical space-time has no any curvature due to an
electromagnetic field. So, the "Weil space" was considered earlier
just as a methodological result having nothing to do with the
reality. But we see that, within the geometrical approach, the Weil
space does not have to be considered as the curved real physical
space-time. It has to be considered as a covering space of the
4-manifold whose proprieties are described by Eq.(12). This space
(being a formal geometrical construction) plays the role of
"internal" space, and gauge transformations (17,18) look now like
natural manifestation of the Eq.(12) invariance about those covering
space transformations that do not change its symmetry (Recall that
the covering space symmetry defines all proprieties of the
above-mentioned 4-manifold).

\begin{center}{\bf 4. Conclusion and final remarks}\end{center}

 The above considerations can be summarized as follows.

1. Equation (12), as well as Eq.(1), can be considered as a
description of some microscopic closed space-time
4-manifold---specific microscopic localized deviation of the space
geometry from the Euclidean one.

2. Gauge transformation of the wave function (16) corresponds to
such coordinates transformation of the manifold covering space that
does not change its symmetry and, therefore, does not change the
mathematical description of the manifold. The manifold covering
space proves to be non-Euclidean "Weil space."

3. Thus, the gauge invariance of Eq.(12) proves not to be some
additional theoretical principle,  but it happens to be a natural
consequence of the basic principles of the proposed geometrical
interpretation. The covering space plays the role of so called
"particle internal space".

As we see, the next steps look as follows. First, it should be a
geometrization of another known gauge groups such as SU(2)-group and
SU(3)-group, and then it should be a geometrical interpretation of
the secondary quantization procedure. The discussion of these
problems will be presented elsewhere.

And the last remark. It is accepted that the absence of hidden
parameters is the necessary condition for validity of any new
quantum mechanics interpretation [16,17]. The suggested geometrical
interpretation does not include any hidden parameters.

\begin{center}\textbf{Appendix}\end{center}

To my knowledge, the topological manifolds were not before
identified by the linear differential equations like Eq.(1). We will
demonstrate, by a simplest example, how linear differential
equations can describe the manifold topological and metric
characteristics.

Let us consider the simplest closed connected manifold, namely, a
one-dimensional manifold $S^{1}$ homeomorfic with (equivalent to)
ring. It is equivalent in the sense that it can be represented as
any of the objects obtained from ring by its deformation without
discontinuities. Different classes of closed paths starting and
ending at the same point of the ring form its fundamental group and
this group is the topological invariant of our manifold. The
fundamental group of $S^{1}$ is the group of integers ($Z$
group)[18]. In turn, $Z$ group is isomorphous with the group of
parallel translations along a straight line with one generator that
we denotes as $l$, and this straight line (x axis) serves as
universal covering space of our manifold (one-dimensional Euclidean
space). So,the universal covering space of the ring is an
one-dimensional Euclidean space with additional symmetry (the
parallel translations group). The topological type of our manifold
does not depend on the concrete value of the parameter $l$. This
parameter is the manifold additional metric (not topological)
characteristic that sets limits on the ring possible configurations.
We will consider $l$ as a given, preassigned quantity, $l=l_{0}$,
and this quantity must be reflected along with the ring fundamental
group within the manifold mathematical description (this restriction
on $l$ is an analogy of restriction (7) that appeared within the
framework of the Dirac equation topological interpretation).

As we noticed in Section 2, the operator
$$T_{x}=\left(\frac{il_{0}}{2\pi}\right)\frac{d}{dx} \eqno (A1)$$
can be considered as the operator representation of a discrete group
of parallel translations along $x$ axis with the generator $l_{0}$.
The basic vectors of this representation are represented by the
functions $\varphi(x)$
$$\varphi(x)=\exp\left(-2\pi i\frac{x}{l_{0}}\right),\eqno (A2)$$
because [7]
$$\varphi^{'}(x^{'})\equiv
T_{x}\varphi(x^{'})=\varphi(x),\quad x^{'}=x+l_{0}.\eqno(A3)$$

Thus, both the ring fundamental group and the metrical restriction
on the length of the parameter of this group are defined by
relationship (A3). Thus, the manifold is fully identified by the
linear differential equation
$$\left(\frac{il_{0}}{2\pi}\right)\frac{d\varphi}{dx}=\varphi(x+l_{0})=\varphi(x),$$ or
$$i\frac{d\varphi}{dx}=m_{0}\varphi,\quad m_{0}=\frac{2\pi}{l_{0}}.\eqno (A4)$$
Therefore, Eq.(A4) can be considered as one of the possible
mathematical description of the ring topological and metrical
properties. Of course, there may be another physical and geometrical
interpretations of Eq.(A4).
\par\bigskip

*E-mail address: olega@gagarinclub.ru

 \noindent 1. O.A. Olkhov, in Proceedings of Int.Workshop on
High Energy
Physics and Field Theory, Protvino, Russia, June 2001, p.327.\\
2. O.A. Olkhov, in Proceedings of "Photon 2001", Ascona,
Switzerland, September 2001, p.360.\\
3. O.A. Olkhov, in Proceedings of Int.Conf. on New Trends in
High-Energy Physics, Yalta (Crimea), September 2001, p.324.\\
4. O.A. Olkhov in Proceedings of "Group 24", Paris, France, July 2002. p. 363.\\
5. O.A. Olkhov, Moscow Institute of Physics and
Technology,Preprint  No.2002-1, Moscow,2002.\\
6. J.D. Bjorken and S.D. Drell, \emph{Relativistic Quantum Mechanics}. (McGraw-Hill Book Company, 1964).\\
7. M. Hamermesh, {\it Group Theory and Its Application to Physical
Problems}. (Argonne National Laboratory, 1964).\\
8. A.I. Achiezer, S.V. Peletminski, \emph{Fields and Fundamental
Ineractions}. (Naukova Dumka, Kiev, 1986), Ch.1\\
9. E. Cartan, \emph{Le\c{c}ons sur la th\'{e}orie des spineurs}.
(Actuallit\'{e}s scientifiques et industrielles, Paris, 1938).\\
10. See, e.g., H.S.M. Coxeter, {\it Introduction to Geometry}. (John Wiley and Sons,Inc., N.Y.-London, 1961).\\
11. G. Kane, {\it Modern Elementary Particle Physics}. (Addison-Wesley Publishing Company, Inc., 1987).\\
12. L.B. Okun, {\it Leptons and Quarks}. (North-Holland Publishing
Co., 1982).\\
13.J.A. Schouten, {\it Tensor Analisis for Physicists}. (Clarendon
Press, Oxford, 1951).\\
14.J.A. Schouten and D.J. Struik, {\it Einf\"{u}hrung in die neue
Methoden der Differentialgeometrie}. (Groningen-Batavia, Noordhoff,
1935), Vol.1.\\
15.H. Weyl, \emph{Gravitation und Electrizit\"{a}t}. (Preuss. Akad. Wiss., Berlin, 1918); Z.f.Phys. \textbf{56}, 330 (1929). \\
16.J.S. Bell, Rev.Mod.Phys. \textbf{38}, 447 (1966)\\
17.J.V. Neumann, \emph{Mathematische grundlagen der
quantenmechanik}. (Springer, Berlin, 1935).\\
18. See, e.g., B.А. Dubrovin, S.P. Novikov, A.T. Fomenko,
\emph{Modern Geometry. Methods and Applications}. (Nauka, Moscow, 1986).\\

\end{document}